# An entanglement-based quantum network based on symmetric dispersive optics quantum key distribution

Xu Liu[1], Xin Yao[1], Rong Xue[1], Heqing Wang[4], Hao Li[4], Zhen Wang[4], Lixing You[4], Xue Feng[1,2], Fang Liu[1,2], Kaiyu Cui[1,2], Yidong Huang[1,2,3] and Wei Zhang[1,2,3, *]

[1]Beijing National Research Center for Information Science and Technology (BNRist), Beijing Innovation Center for Future Chips, Electronic Engineering Department, Tsinghua University, Beijing 100084, China

[2] Frontier Science Center for Quantum Information, Beijing 100084, China

[3]Beijing Academy of Quantum Information Sciences, Beijing 100193, China

[4]State Key Laboratory of Functional Materials for Informatics, Shanghai Institute of Microsystem and Information Technology, Chinese Academy of Sciences, Shanghai 200050, China

* zwei@tsinghua.edu.cn

**Quantum key distribution (QKD) is a crucial technology for information security in the future. Developing simple and efficient ways to establish QKD among multiple users are important to extend the applications of QKD in communication networks. Herein, we proposed a scheme of symmetric dispersive optics QKD (DO-QKD) and demonstrated an entanglement-based quantum network based on it. In the experiment, a broadband entanglement photon pair source was shared by end users via wavelength and space division multiplexing. The wide spectrum of generated entangled photon pairs was divided into 16 combinations of frequency-conjugate channels. Photon pairs in each channel combination supported a fully-connected subnet with 8 users by a passive beam splitter. Eventually, it showed that an entanglement-based QKD network over 100 users could be supported by one entangled photon pair source in this architecture. It has great potential on applications of local quantum networks with large user number.**

Information security is of significant importance in many applications. Nowadays, the security of these applications is mainly based on public-key cryptography[1,2] which assumes that the computation power is limited. Quantum key distribution (QKD) can provide cryptographic keys with information-theoretic security by the nature of quantum physics[3-6], which may revolutionize the protection way of information exchange in the future. Since the first protocol[7] (BB84) was proposed, QKD has been developed significantly[8-15]. Many field tests of QKD has been implemented, proving it to be a reliable technology[16-18].

In recent years, how to build quantum networks conveniently and efficiently becomes the focus of research on QKD. The quantum nodal network has achieved great developments[19-21], which connects multiple point-to-point QKD links by trusted nodes.

In the future, it may be also fulfilled by the quantum relayed satellite[22-24]. Although it is promising on realizing long distance backbone QKD networks, it is not efficient to provide QKD services in local networks with many users and complicated connections, such as networks in companies, campuses and communities, etc. To improve the efficiency of QKD networks, the concept of quantum access network is proposed, which allows multiple users share the receivers or sources[25]. The first quantum access network was proposed by Townsend[26], in which single photons from a central node are distributed to multiple end users by a passive beam splitter. It realizes point-to-point QKD between the central node and each end user, achieving a point-to-multipoint network. The secure connection between these end users depends on the relay of the central node. Hence, the security of the whole network extremely relies on the fidelity of the central node.

Quantum entanglement is the crucial resource for certified generation of shared randomness[27,28], quantum communication[29,30] and so on. The flexibility on entanglement distribution among multiple users may provide new ways to realize QKD networks[31]. In a previous work, the signal and idler photons from entangled photons were divided and distributed to two sets of users, respectively[32]. In each side, an optical switch was used to distribute photons to a specific user. By this way, entanglement-based QKD can be established between the two specific users at the two sides, achieving a quantum network with active routing function. However, the users in the same side cannot establish QKD in this architecture. What's more, the network efficiency would be limited to some duty cycle of optical switches, which is a common problem for the point-to-multipoint architectures based on optical switches[33]. Recently, a fully-connected QKD network was proposed and demonstrated experimentally based on a broadband polarization-entangled quantum light source[34]. Each two users in the network were allocated with photon pairs in two correlated wavelength channels by wavelength division multiplexing technology. It can be expected that a minimum of $N\times(N-1)$ wavelength channels are required to fully connect $N$ users in this architecture. Hence, it rapidly depletes the resource of quantum light source bandwidth as the user number increases.

In this paper, we develop a quantum network based on another way of quantum entanglement distribution, in which the entangled photon pairs generated by a quantum light source are sent to $N$ end users by a 1×$N$ beam splitter directly. In this way, the two photons in each pair may be distributed to any users randomly. Hence, any end users will have coincidence events, which can be used to establish a QKD network with fully connection conveniently. However, it can be expected that the rates of coincidence events between users will reduce rapidly if the user number supported by the quantum light source increases. To utilize the coincidence events efficiently, we use the dispersive optics QKD (DO-QKD) based on energy-time entanglement to achieve the QKD network. In the scheme of entanglement-based DO-QKD, signal and idler photons are sent to two users. Normal and anomalous dispersion components are introduced at the two sides to carry out the security test, which is guaranteed by the nonlocal dispersion cancellation effect[35] of energy-time entangled photon pairs. It has been proven to be secure against collective attacks[36,37]. An attractive property of the entanglement-based DO-QKD is that high dimensional time encoding can be utilized in this scheme, which supports multi-bit key generation per coincidence[38], improving the utilization efficiency of coincidence events.

However, as shown in Fig. 1 (a), the two users have different configurations in conventional entanglement-based DO-QKD schemes where one user has normal dispersion component and the other has anomalous dispersion component. It cannot be used to establish the QKD network based on a 1×$N$ beam splitter. Therefore, we propose a modified scheme named as "symmetric DO-QKD", which is shown in Fig.1 (b). In the proposed scheme, the two users have both normal and anomalous dispersion components. The paths with two different dispersion modules in the two users are treated as the measurement bases. There are two bases between the two users and both of them experience the effect of nonlocal dispersion cancellation. They are used for security test and key generation, respectively, which should be predefined between any two users according to the nonlocal dispersion cancellation requirement. It is named "symmetric" since the users have the same configurations, which solves the problem of the conventional one. Hence, the symmetric DO-QKD scheme can be used in the network based on a 1×$N$ beam splitter.

To further improve the utilization efficiency of the entanglement resource provided by the quantum light source, we also introduce wavelength multiplexing in this entanglement-based QKD network, exploring how many users can be supported by one quantum light source in this way. Fig. 1(c) shows the sketch of the network architecture. The quantum light source located in the central node generates energy-time entangled photon pairs over a wide wavelength region, which are divided into different channels by wavelength division multiplexing. The signal and idler photons in two frequency-conjugate wavelength channels are entangled and contribute to coincidence events. They are selected and multiplexed into one optical fiber, then distributed to $N$ users by a 1×$N$ beam splitter. Symmetric DO-QKD can be established between any two of these $N$ users, realizing a fully-connected QKD subnet. The beam splitter also can be placed in the central node in the scenario of local networks. If the quantum light source could support *M* frequency-conjugate wavelength channel combinations, it can support *M* QKD subnets. To connect these subnets, the simplest way is that each subnet provides a user to the central node, as shown in Fig. 1(c). In this architecture, the central node acts as a trusted relay node. Any two users in different subnets can eventually establish cryptographic keys through the central node. The logical topology structure of this architecture is shown in Fig. 1(d). Each subnet has a fully connected mesh topology, while all the subnets are connected by the central node.

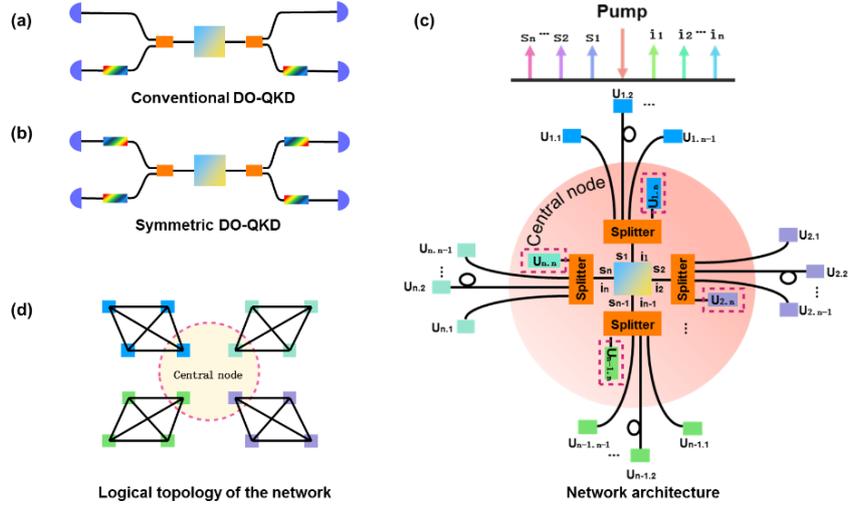

Fig. 1 Illustration of the proposed QKD network. (a) The sketch of the conventional DO-QKD and (b) that of the symmetric DO-QKD. (c) The sketch of the network architecture. The illustration on the top shows that the generated photon pairs of the quantum light source are over a wide wavelength region, which are divided into different wavelength channels by wavelength division multiplexing. The signal and idler photons in two frequency-conjugate wavelength channels is multiplexed together and further distributed by a 1×*N* beam splitter to *N* users, supporting a subnet of QKD. In the scenario of local networks, the quantum light source and the 1×*N* beam splitters are set in the central node. Each subnet also provides a user to the center node which are shown by the dash square in the figure. Two users in different subnets can eventually establish cryptographic keys through the central node. (d) The logical topology of the network. Each subnet is a fully-connected QKD network, and provides a user to the central node which connects all the subnets.

It can be seen that this QKD network architecture utilizes the entanglement resource with high efficiency and robustness. Firstly, the symmetric DO-QKD with high dimensional time encoding is applied to enhance the utilization efficiency of the coincidence events. Secondly, the broadband characteristics of quantum light source is utilized sufficiently by wavelength division multiplexing to support many subnets. Finally, if the user number of each subnet is not too small, most photon pairs would be distributed to two different users randomly, which is a simple but efficient way to realize a fully-connected subnet. On the other hand, only the trust on the central node is required. The insecurity of an end user will not impact the QKD between other users in a subnet. If the central node is insecure, the cryptographic keys between users in different subnets will be insecure. However, it will not impact the QKDs between users in the same subnet.

**Results**

**Experimental system.** To demonstrate the proposed QKD network architecture based on symmetric DO-QKD, we establish the experimental system shown in Fig. 2.

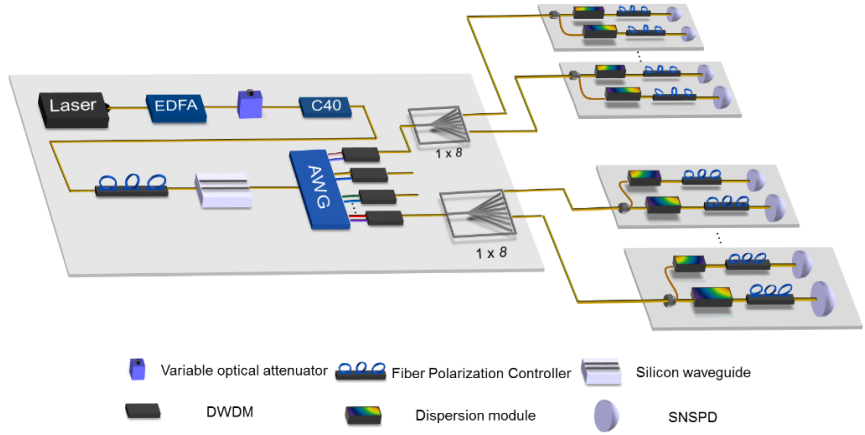

Fig. 2 Experimental system of the QKD network based on symmetric DO-QKD. In each user, the normal and anomalous dispersion modules are introduced for the nonlocal dispersion cancellations. The photons are detected by superconducting nanowire single-photon detectors (SNSPDs). This experimental system supports 16 subnets, and each subnet has 8 end users. DWDM: dense wavelength division multiplexing component; ND (AD): normal dispersion (anomalous dispersion) components; FPC: fiber polarization controller.

The energy-time entangled photon pairs are generated in a broad telecom band through the spontaneous four wave mixing (SFWM) effect in a piece of silicon waveguide under continuous wave (CW) pumping. The silicon waveguide was 3 mm in length on a silicon-on-insulator (SOI) photonic chip. An arrayed waveguide grating (AWG) is used to filter the output photons. In the experimental system, the wide spectrum of the signal and idler photons is divided into 32 wavelength channels. Each combination of correlated wavelength channels that satisfy the energy conservation condition of SFWM are further multiplexed together by a dense wavelength division multiplexing (DWDM) component and then distributed randomly to 8 users by a 1×8 planar lightwave circuit splitter (PLCS). Hence, the source supports 16 subnets. In each end user, the photons are split into two paths by a fiber coupler and detected by superconducting nanowire single photon detectors (SNSPDs), respectively. The fiber polarization controllers (FPCs) before the SNSPDs are used to maximize the detection efficiencies. A normal dispersion component and an anomalous dispersion component are introduced in the two paths, respectively. Hence, all the end users have the same configurations and any two users in a subnet can establish the symmetric DO-QKD and generate cryptographic keys between them. The cryptographic keys between users in different subnets are relied on the central node, which includes the quantum light source and the PLCSs. In each subnet, one specific user is placed in the central node for the connections between different subnets. Considering the scenario of local networks, the fiber lengths between the central node and end users are tens of meters in the experiment system. Hence, the impact of fiber dispersion can be neglected in the experiments. See the experimental details in Supplement 1.

**Entanglement distribution.** High-quality entanglement distribution is crucial in many fields like QKD[4,29], quantum teleportation[39,40] and so on. It was firstly tested in the experimental system shown in Fig. 2. First of all, we tested the broadband characteristics of the quantum light source. The AWG used in the experiment covers the International Telecommunication Union (ITU) channels of C21-C60. To fully utilize the filtering channels of the AWG, the mono-color pump light of the quantum light source is set at 1545.32 nm, the central wavelength of the channel of C40. The correlated wavelength channels for signal and idler photons are distributed symmetrically around this channel of AWG. We selected channels of C44~C59 as the signal channels and channels of C21~C36 as the idler channels. In the measurement, the SNSPDs (with FPCs) were connected to the output fiber of the AWG for specific ITU channels. The performances of the wavelength division in the system and corresponding coincidence results are shown in Fig. 3. The single photon count rates of these channels were measured under a specific pumping level which was fixed in the following experiments, which are shown in Fig. 3(a). It can be seen that the photon count rates of all the channels are close due to the broadband characteristics of the photon pair generation by SFWM in the silicon waveguide. The coincidence counts of each combination of correlated signal and idler channels are shown in Fig. 3(b). For clarity, the coincidence peak positions of different channel combinations were shifted with a fixed time delay of 1600 ps. It can be seen that signal and idler photons in all the correlated channel combinations have good coincidence. Each channel combination can be used to constitute a subnet. Therefore, 16 subnets can be supported by the quantum light source in the experimental system.

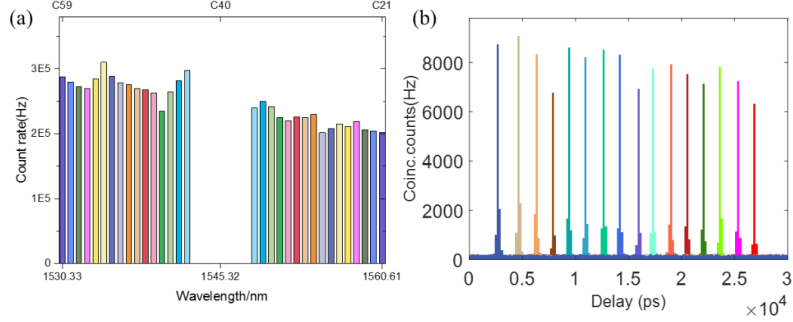

Fig. 3 The performances of the wavelength division in the system and corresponding coincidence results. (a) Single photon count rates of different wavelength channels. (b) Experimental results of coincidence counts of 16 correlated wavelength channel combinations under a time bin width of 192 ps.

In the experiment system, the signal and idler photons with channels of C31 and C49 are multiplexed and distributed to 8 end users of a subnet by a 1×8 PLCS. By this way, each two users in the subnet have coincidence events of entangled photon pairs, which is the base to realize the symmetric DO-QKD network. To show the performance of the entanglement distribution between the end users, all the coincidence events of 28 user combinations in this typical subnet are measured by the SNSPDs. The results are shown in Fig. 4, in which the two figures show the measured coincidence count rates and the corresponding CARs between any two users, respectively. It can be seen that any two users in the subnet network have coincidence events from the entangled photon pairs. They support the fully-connected subnet, even though the distribution by the 1×8 PLCS reduces the coincidence rates to several tens of counts per second. On the other hand, all the CARs are higher than 100, indicating that the high quality of the entanglement maintains after its distribution, which could support the high performance QKD.

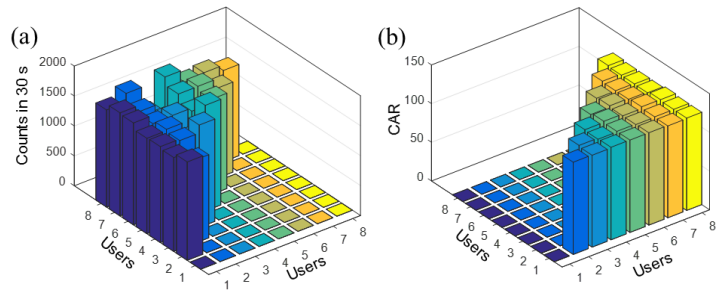

Fig. 4. Results of entanglement distribution of all the user combinations in a typical subnet acquired in 30s under a time bin width of 192ps. (a) The coincidence counts. (b) The CARs.

**Key generation in the network.** In the experiment system, the entanglement-based symmetric DO-QKD protocol was used to generate cryptographic keys between the end users in the network. The high dimensional time encoding was utilized in the symmetric DO-QKD scheme, which supported multi-bit key generation per coincidence (See in Supplementary Materials), improving the utilization efficiency of the entanglement resource in this network. As shown in Fig. 2, both normal and anomalous dispersion components are introduced in each user in this scheme. Thus, all the end users in the network have the same configurations. For any two users, there are two bases between them and both of them experience the effect of nonlocal dispersion cancellation using the matched dispersion components. The coincidences under the two bases are used for the security test and key generation, which are named as "S base" and "K base", respectively. If one user uses the path with the normal dispersion component as the S base, the other user should select the path with anomalous dispersion component as the S base, and vice versa. Then the remaining measurement base is used as the K base. The bases selection between the users in the subnet should be predefined before QKD operation.

We measured the performances of entanglement-based symmetric DO-QKDs in the experimental system. Typical results of coincidence counts under four possible measurement base combinations between two end users in a subnet (supported by photons with channels of C31 and C49) are shown in Fig. 5. In these figures, K1, K2, S1, S2 indicate the measurement bases predefined between two end users. It can be seen that the coincidence peak is narrow due to the nonlocal dispersion cancellations if it is measured under K bases at both sides, which are shown in Fig. 5 (a). It is also narrow if it is measured under S bases at both sides, which are shown in Fig. 5 (d). However, the coincidence peaks are broadened if they are measured under different bases, which are shown in Fig. 5(b) and (c). These results show that the symmetric configurations in these users are able to realize DO-QKD. According to the single photon detection events measured under K1-K2 base, the performance of raw key generation between these two users can be optimized by adjusting the parameters of the high dimensional time encoding (Please See in Supplementary Materials). Eventually, a raw key rate of 80.9 bits per second (bps) can be achieved with a quantum bit error rate (QBER) lower than 5% under an optimized time encoding format, which generates 4 bits of raw keys per coincidence event. On the other hand, the secure information that two users could extract per coincidence also can be calculated according to the experimental results shown in Fig. 5, by which the secure key rate between the two users after privacy amplification can be estimated to 63.7 bps (See Supplementary Materials for details of the analysis).

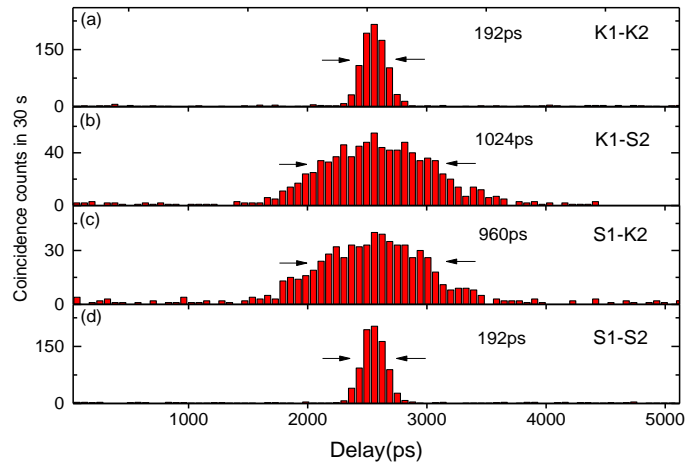

Fig. 5 A typical result of coincidence counts between two end users under four possible measurement basis combinations acquired in 30 s. (a) Coincidence counts under K1 and K2. (b) Coincidence counts under K1 and S2. (c) Coincidence counts under S1 and K2. (d) Coincidence counts under S1 and S2.

Further, we measured the performances of symmetric DO-QKDs of all the user combinations in this subnet. The results are shown in Fig. 6, in which the blue and yellow columns indicate the generation rates of raw keys and secure keys, respectively. Their difference is due to the costs of error correction and privacy amplification. It can be seen that any two users in the subnet can generate cryptographic keys by the symmetric DO-QKD, with an average secure key rate of ~60 bps.

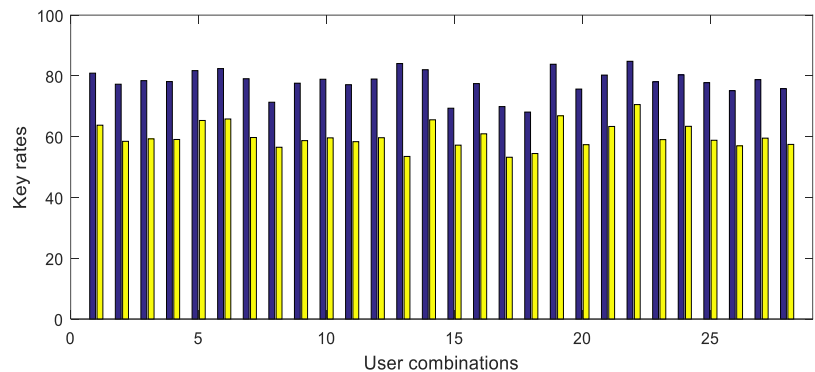

Fig. 6 The performances of symmetric DO-QKDs of all the user combinations in the subnet. The blue and yellow columns indicate the generation rates of raw keys and secure keys between each two users, respectively.

In the network architecture shown in Fig. 1(c), cryptographic key generation between users in different subnets is relied on the specific users in the central node. The two users in different subnets firstly generated keys with the corresponding users of their subnets in the central node by symmetric DO-QKD. Then the central node performed a bitwise exclusive OR operation between the two keys and sent the new key via a classical channel to one of the users. Eventually, the user can decode the other one's original key by another bitwise exclusive OR operation, by which the two users share the same cryptographic keys. Eventually, a network with 112 users (16 subnets with 7 end users per subnet) can be realized in the experimental system based on symmetric DO-QKD and the architecture shown in Fig. 1.

**Discussion**

The proposed symmetric DO-QKD has two effects in this network. First, the users have the same configurations in this scheme, which make it feasible to introduce the DO-QKD into the network based on entanglement distribution through a $1 \times N$ beam splitter. Second, the high dimensional time encoding used in symmetric DO-QKD is beneficial to improve the utilization efficiency of the coincidence events that are precious resources for QKD in such a network architecture with large number of users.

In this experiment, the fiber lengths between the central node and the end users are short under the consideration of the scenario of local networks. It can be expected that the geographical scale of this network architecture could be extended by introducing long distance fiber transmissions with proper compensations for the fiber dispersions and fine clock distribution for time synchronization. It is also worth noting that wavelength multiplexing technology of entanglement resources provided by a broadband quantum light sources is quite flexible in constituting quantum networks[31,41]. It would further bloom more efficient and practical network architectures based on the proposed symmetric DO-QKD for other interesting application scenarios.

**Funding.**

National Key R&D Program of China (2017YFA0303704, 2018YFB2200400 2017YFA0304000); National Natural Science Foundation of China (NSFC) (61575102, 61875101, 91750206, 61621064); Beijing National Science Foundation (BNSF) (Z180012); Beijing Academy of Quantum Information Sciences (Y18G26).

**Disclosures**

The authors declare no conflicts of interest.